\documentclass[prl,twocolumn,amsmath,amssymb,superscriptaddress]{revtex4}
\usepackage{graphicx}

\begin{document}

\title{Manifestation of the spin-Hall effect through transport measurements in the mesoscopic regime}
\author{E. M. Hankiewicz}
\affiliation{Department of Physics, Texas A\&M University, College
Station, TX 77843-4242}
\author{L.W. Molenkamp}
\affiliation{Physikalisches Institut (EP3), Universit\"{a}t W\"{u}rzburg, Am Hubland, D-97074 W\"{u}rzburg, Germany}
\author{T. Jungwirth}
\affiliation{Institute of Physics ASCR, Cukrovarnick\'a 10, 162 53
Praha 6, Czech Republic} \affiliation{School of Physics and
Astronomy, University of Nottingham, Nottingham NG7 2RD, UK}
\author{Jairo Sinova}
\affiliation{Department of Physics, Texas A\&M University, College
Station, TX 77843-4242}

\date{\today}

\begin{abstract}
We study theoretically the manifestation of the spin-Hall effect in a
two-dimensional electronic system with Rashba spin-orbit
coupling via dc-transport measurements in realistic mesoscopic
H-shape structures. The Landauer-Buttiker formalism is used to model
samples with mobilities and Rashba coupling strengths
of current experiments and to demonstrate the appearance of a measurable
Rashba-coupling dependent voltage. This type of measurement
requires only metal contacts, i.e., no magnetic elements are present.
We also confirm  the robustness of the intrinsic spin-Hall effect
against disorder in the mesoscopic metallic regime in agreement with results
of exact diagonalization studies in the bulk.
\end{abstract}

\maketitle

{\it  Introduction.}  The ability  to manipulate  electronically spins
and to generate  spin currents in semiconductors is  the {\em sine qua
non}  for  the full  development  of  semiconductor based  spintronics
\cite{Wolf01}.
The  control  of  spin  and spin-currents  without  applying  external
magnetic fields can be  achieved through the spin-orbit (SO) coupling,
which acts  as an  effective momentum-dependent Zeeman  field.  Within
this context, the newly discovered intrinsic spin-Hall effect (SHE) in
p-doped semicondcutors  by Murakami {\it  et al}.\cite{Murakami03} and
in a  two-dimensional electron  system (2DES) by  Sinova {\it  et al.}
\cite{sinova04} offers new  possibilites for  spin current manipulation
and  generation in high  mobility paramagnetic  semiconductor systems.
The  intrinsic  spin-Hall effect  represents  a spin-current  response
generated perpendicular  to the driving electric field.  The spins are
tilted out of the plane due  to the torque imparted by the SO coupling
induced effective  Zeeman field.  In  the Rashba SO  coupled 2DESs  the bulk
intrinsic spin-Hall conductivity was found to have a value of $e/8\pi$
in  the clean limit  for the  case of  both spin-split  subbands being
occupied and  decreases linearly with the electron  density for single
spin-split subband occupation \cite{sinova04}.

The SHE  discovery has generated  a tremendous  interest in  the research
community
\cite{Schlie03,All_part,Rashba03,Rashba04,Dimitrova04,Halperin04,
kentaro04,Inoue04, Chalaev04,Khaetskii04,Xie04,Nikolic04,Sheng04}.
Similarly to the  long-standing debate on the origin  of the anomalous
Hall  effect  (AHE) \cite{reviewJairo},  the  robustness  of the  bulk
intrinsic SHE against disorder and  how it is related to the scattering
mediated extrinsic spin-Hall effect \cite{Perel,Hirsch99,Zhang00}, has
been    the     focus    of    an     intense    theoretical    debate
\cite{Schlie03,Dimitrova04,Halperin04,kentaro04,Inoue04,Chalaev04,Khaetskii04}.
While it  was understood  originally \cite{sinova04} that,  unlike the
quantum Hall effect,  the universal value of the  intrinsic SHE in the
Rasbha  SO  coupled 2DESs  will  be  reduced  whenever the  disorder
broadening is larger  or similar to the SO  coupling splitting, as was
verified    within    a    standard    Born-approximation    treatment
\cite{Schlie03},  taking  into  account the  ladder  vertex
corrections through  various methods suggests that  the bulk spin-Hall
conductivity    vanishes    in     the    weak    disorder    dc-limit
\cite{Inoue04,Halperin04}.    However,   these   results   have   been
challenged by other analytical calculations which also consider ladder
vertex    corrections   \cite{Chalaev04}.    Other    recent   studies
\cite{Khaetskii04}, using arguments that echo the long-standing debate
between skew and side-jump scattering in the AHE, have argued that the
intrinsic SHE vanishes in all regimes.

Given  the  ferraginous collection  of  analytical results,  unbiased
numerical calculations  are needed to shed light  on this controversy.
An  exact diagonalization treatment  of disorder  \cite{kentaro04} has
shown that the bulk intrinsic  SHE is robust against weak disorder. In
addition,  several numerical  studies  utilizing the  Landauer-Buttiker
(LB) formalism  in a tight-binding  model representation  of the  Rashba SO
coupling  Hamiltonian  in  the   presence  of  disorder  show  similar
conclusion in  the limit which corresponds to  the continuum effective
mass model (see below) \cite{Datta,Xie04,Nikolic04,Sheng04}.

\begin{figure}[h]
\includegraphics[width=3.25in]{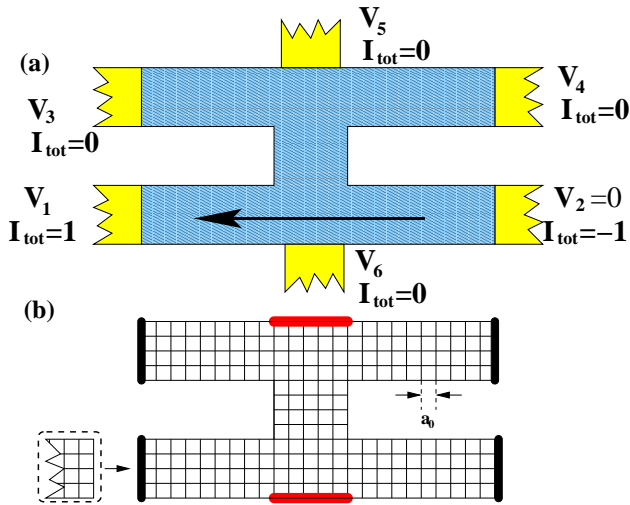}
\caption{(a) Mesoscopic H-bar probe with metallic leads. (b) The continuum model
is converted into a tight-binding model and the effects of the leads are treated exactly
through a self-energy.}
\label{fig1}
\end{figure}

In this paper we address a  key question that has yet to be addressed
directly: {\it  How to measure  the intrinsic
spin-Hall  effect  through   transport  measurements?}   All  previous
numerical  studies  have  focused  on the  controversy  regarding  the
robustness  of the effect  against disorder  and how  disorder, Rashba
coupling  strength, etc.,  change the  continuum effective  mass model
value of $e/8\pi$  in the various models.  Recently  a new
theoretical question has
arisen,  whether  the   dissipationless  currents  or  spin  background
currents can lead to spin accumulation or to a steady signal which can
manifest such  an effect \cite{Rashba03,Rashba04}.   This question can
be  addressed unambiguously  in  the mesoscopic  regime where  the
effect of  the leads  and disorder can  be taken into  account
through  the  explicit  treatment of  voltage  and current  probes
within the  LB formalism \cite{Datta,Xie04,Nikolic04,Sheng04}. Here we
consider an H-shape structure  shown in Fig.~\ref{fig1} to demonstrate
the  appearance  of  the  spin-Hall effect  through  dc-transport
measurements without any magnetic elements.  A current is
driven from lead 2 to 1  in the lower leg shown in Fig.~\ref{fig1} (a)
and the rest of the contacts  are voltage probes, i.e., leads with total
zero current.  Then, for typical  current utilized in  experiments and
typical parameters  we find a voltage  difference $\Delta V_{34}\equiv
V_3-V_4$ dependent on the Rashba coupling and coupled to the spin-Hall
conductance defined at, e.g., probe  6. The variation of this voltage,
relative to the  residual voltage obtained at  zero Rashba coupling,  is of the
order  $\mu V$  for the  maximum  system sizes  that we  can model
($\sim 0.1\mu$m).

In  the mesoscopic  regime,  several of  the  controversies that
have arisen from the  study of the bulk spin-transport
coefficients can be addressed.  Within  this regime the only
assumption  made in describing the transport  through the  sample
via  the current and voltage probes, other  than the applicability
of the tight-binding approximation for the SO coupled electronic
structure, is  that such contacts are perfectly metallic, i.e., an
exact analytical expression is known for their Green's functions.

Several  numerical studies have  addressed  the robustness  of
the  intrinsic spin-Hall   effect  within   the  mesoscopic
regime   utilizing  the LB formalism in a tight-binding model
representation of the          Rashba          SO          coupled
Hamiltonian \cite{Datta,Xie04,Nikolic04,Sheng04}.   Xie {\it et
al}. \cite{Xie04} obtained  the expected  universal  spin-Hall
conductance  in the  weak disorder  limit with  the  Fermi
energy, $E_F$,  at  the band  center $E_F=0$,  but observed  that
spin-Hall  conductance  decreased rapidly with system  size due
to localization effects.  These results  are in direct
contradiction  to the general  notion that the SO  coupled 2DESs
exhibit   a  delocalized  region   \cite{Sheng04,Ando89}  and,
perhaps  more importantly, that at the band center of this model
any Hall coefficient vanishes due to electron-hole symmetry
\cite{Nikolic04,Sheng04}. Perhaps the most compendious of
these tight-binding model numerical studies are Refs. \onlinecite{Nikolic04}
and \onlinecite{Sheng04} where the expected   symmetry  of  the   Hall  conductance   
with respect to $E_F$   and  the expected metal-insulator transition as a
function of SO coupling and disorder strength  \cite{Ando89}.

{\it  Model   Hamiltonian  and  LB   treatment  of  the spin-Hall
effect}.  The experimental detection of the SHE through electrical
means is conceptually challenging. Given  the   controversy
surrounding  the   nature  of   the spin-currents  generated by
electric fields, a measurement of the voltage  between two
metallic contacts \cite{Rashba04} appears to be the most promising
dc-transport approach to unambiguously determine the presence of
the SHE signal.  We focus our attention then in  the proposed
H-shape device shown in Fig.~\ref{fig1}  and demonstrate that
within the  mesoscopic metallic regime  the intrinsic  SHE is
exhibited through  the change in a voltage difference between two
contacts as the Rashba spin-orbit coupling  is varied.

The   continuum effective  mass  model  described   by  the  2DES Hamiltonian with
the Rashba SO interactions is given by
$\hat{H}=\frac{\hat{p}^2}{2m^*}+\lambda(\hat{\sigma}_xp_y-\hat{\sigma}_yp_x)
+H_{dis}$,
where    the    second    term    is    the    Rashba    SO    coupling
\cite{Rashba60,Rashba84}  due  to   the  asymmetry  of  the  confining
potential and  $H_{dis}$ is the disorder potential
To model  the complex geometry of our disordered  conductor  within  the  LB
formalism  we use the tight-binding  (or finite
differences)  approximation \cite{Datta,Xie04}.  
Within this approximation the continuum  effective mass  envelope  function
Hamiltonian becomes: 
\begin{eqnarray}
H&=&\sum_{j,\sigma} \epsilon_j c^{\dag}_{j,\sigma} c_{j,\sigma}
-t\sum_{j,\vec{\delta},\sigma}  c^{\dag}_{j+\vec{\delta},\sigma} c_{j,\sigma}\nonumber \\
&+&t_{SO}[\sum_{j}-i(c^{\dag}_{j,\uparrow} c_{j+a_{y},\downarrow}
+c^{\dag}_{j,\downarrow} c_{j+a_{y},\uparrow})\nonumber \\
&+&\sum_{j}(c^{\dag}_{j,\uparrow} c_{j+a_{x},\downarrow}
-c^{\dag}_{j,\downarrow} c_{j+a_{x},\uparrow})+h.c.],
\end{eqnarray}
where  $t=\hbar^2/2m^*a_0^2$ and  $t_{SO}=\lambda/2a_0$,  $a_0$  is the
mesh lattice  spacing, and $\vec{\delta}=\pm a_0 \hat{x},\pm a_0 \hat{y}$.
The first term represents  a quenched disorder
potential and disorder is introduced by randomly selecting the on-site
energy $\epsilon_j$ in the range  [-W/2,W/2].  Within the leads the SO
coupling  is zero  and therefore  each  lead should  be considered  as
having   two  independent   spin-channels.   These  leads   constitute
reservoirs  of  electrons  at chemical  potential  $\mu_1,\dots,\mu_N$
where $N$ is  the number of leads which we consider  to be either four
(lead 1-4 in Fig.~\ref{fig1}) or six (leads 1-6 in Fig.~\ref{fig1}).

In the  low temperature limit  $k_B T<<E_F$ and for  low bias-voltage,
the {\it particle}
current going through a particular channel is given within the LB formalism
by \cite{Datta}
$I_{p,\sigma}=(e/h)\sum_{q\sigma'}T_{p,\sigma;q,\sigma'}[V_p-V_q]$,
where  $p$  labels  the   lead  and  $T_{p,\sigma;q,\sigma'}$  is  the
transmission  coefficient  at  the  Fermi  energy  $E_F$  between  the
$(p,\sigma)$ channel and  the $(q,\sigma')$ channel. This transmission
coefficient     is     obtained    by     $T_{p,\sigma;q,\sigma'}={\rm
Tr}[\Gamma_{p,\sigma}G^R\Gamma_{q,\sigma'}G^A]$                    where
$\Gamma_{p,\sigma}$ is given by
$\Gamma_{p,\sigma}(i,j)=i[\Sigma^R_{p,\sigma}(i,j)-\Sigma^A_{p,\sigma}(i,j)]$,
and the retarded and advance  Green's function of the sample $G^{R/A}$
with   the  leads  taken   into  account   through  the   self  energy
$\Sigma^{R/A}_{p,\sigma}(i,j)$ is given by
\begin{equation}
G^{R/A}(i,j)=[E\delta_{i,j}-H_{i,j}-\sum_{p,\sigma}\Sigma^{R/A}_{p,\sigma}(i,j)]^{-1}.
\end{equation}
Here the position  representation of the matrices $\Gamma_{p,\sigma}$,
$G^R$, $H_{i,j}$, and $\Sigma^R$ are in the  subspace of the sample, i.e., it
only  includes sites  in the  sample.
Since the SO coupled Hamiltonian preserves time reversal symmetry, the transmission coefficients obey the relation
$T_{p,\sigma;q,\sigma'}=T_{q,-\sigma';p,-\sigma}$  \cite{Pareek04}.
\begin{figure}[h]
\includegraphics[width=3.25in]{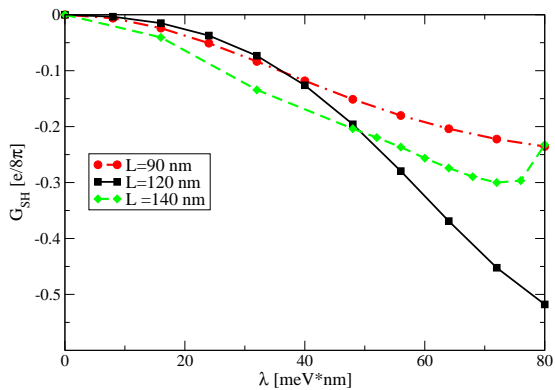}
\caption{Spin-Hall conductance defined at lead 6 (shown in
Fig.~\ref{fig1}) vs. spin-orbit coupling strength for different
size systems for $m_0 =0.05m_e$,$\mu =250,000 cm^2/Vs$ and flowing
current of 10nA in the bottom leg. Here $L$ is divided in 42
points. Only a few disorder realizations are needed for
convergence in samples with these mobilities.}
\end{figure}

Within the  above formalism the  spin current through each  channel is
given        by        $I^s_{p,\sigma}=        (e/4\pi)\sum_{q\sigma'}
T_{p,\sigma;q,\sigma'}[V_p-V_q]$   and  through   this  we   define  a
spin-Hall conductance as
\begin{equation}\label{gspin}
G_{SH}=\frac{(I^s_{6\uparrow}-I^s_{6\downarrow})}{V_1-V_2},
\end{equation}
as  indicated in Fig.~\ref{fig1}.
All the  voltages are  obtained by
imposing  the  boundary conditions  $I_{i,\uparrow}+I_{i\downarrow}=0$
for     i=3    through     6,     $I_{1,\uparrow}+I_{1\downarrow}=1$    and
$I_{2,\uparrow}+I_{2\downarrow}=-1$. The arbitrary  zero of voltage is
fixed by  setting $V_2=0$. These  are later translated to  a realistic
voltages by setting the current to a typical value of $10$ nA.

{\it Results  and discussion.}  In order  to address the  key issue of
how  the spin-Hall  effect  can manifest  itself  through a  dc-transport
measurement  without  ferromagnetic   contacts  we  choose  realistic
parameters  for  our  calculations  which  model  currently  available
systems \cite{Laurens}.   We consider an effective  mass of
$m^*  =0.05m_e$  and a disorder  strength  of  $W=0.09$  meV
corresponding  to  the mobility of  250,000  cm$^2$/Vs,  which is typical  for
a semiconductor like (In,Ga)As.  We take the  Rashba parameter $\lambda$
in the range from  0  to 80  meV~nm  which are  easily obtained  in  experiments
\cite{Koga02,Johnston02},   and   we   choose   the   electron
concentration of  approximately $n_{2D}=10^{12}$ cm$^{-2}$.  The Fermi
energy  is obtained from  the chosen  electron concentration  assuming
an infinite 2DES.  This gives a small difference of a few percent
when  considering  our  finite   tight-binding  model  but  the  leads
themselves  are the  ones  providing the  reservoir  of electrons  and
therefore such fluctuations are  small as verified by direct numerical
calculations (not shown).

In Fig.2 we  show the spin Hall conductance $G_{SH}$  as a
function of Rashba parameter for the H-shape structure  when
current flows through the bottom leg as indicated in
Fig.~\ref{fig1}. Here we consider  the system with 6 leads and
with leg lengths L varying from  90 to 140 nm. The total size  of
the system is $L$  in  the $x$-direction  and  $L/2$ in  the
$y$-direction.  The  horizontal connection bar is $L/6$ by $L/6$.
The width of the legs is $L/6$ with the attached leads of the same
width. These ratios were chosen for typical fabricated samples (of
larger system size) but any  shape is feasible to do and a search
for an optimal geometry is underway \cite{Daumer}.
\begin{figure}[tbh]
\includegraphics[width=3.35in]{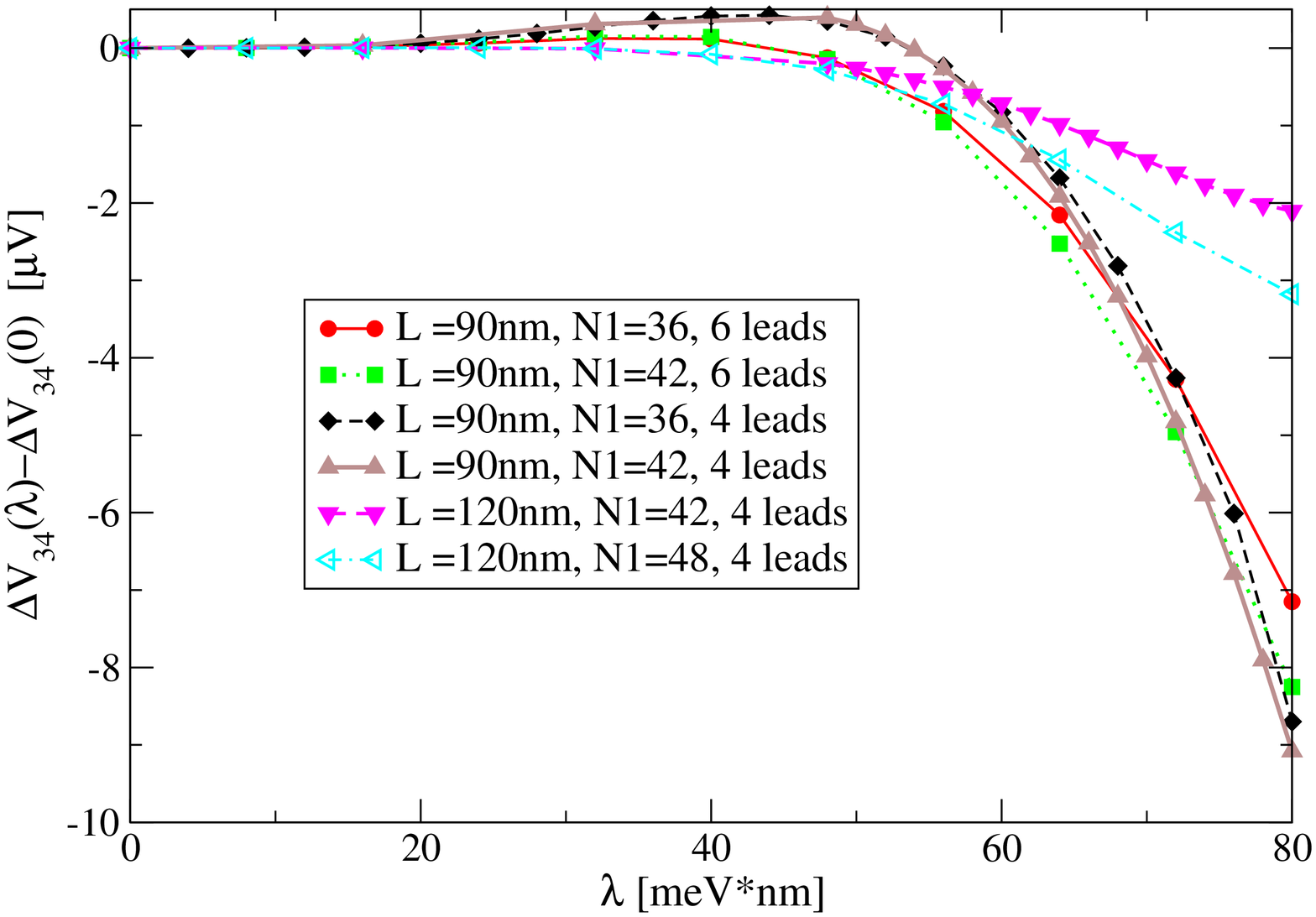}
\caption{The voltage difference between leads 3 and 4 as a
function of Rashba coupling for H-probe  for different size
systems and meshes for $m_0 =0.05m_e$, $\mu=250,000 cm^2/Vs$ and flowing
current of 10nA in the bottom leg.}
\end{figure}

This  particular H-shape  structure allows  for minimalization  of the
residual voltage drop due to charge  flow which is of the order of few
hundreds  of nanovolts.   $G_{SH}$ is  calculated accordingly  to Eq.~
\ref{gspin}. The calculations are conducted for a few different meshes
$N_1=L/a_0$ to  check the  convergence of  results for  $a_0 \rightarrow
0$.  The magnitude of $G_{SH}$ is around 0.2-0.6  in $e/8 \pi $
units for Rashba coupling 70-80 meV~nm and  $L=120-140$ nm.
We also note that within these parameters we are well within the
metallic regime \cite{Sheng04,Ando89} and both spin-split subbands are occupied.

The spin Hall conductance cannot be measured by the paramagnetic
leads and  ferromagnetic  leads  can introduce    spurious effects coming from the
impedance mistmatch preventing ballistic contacts
\cite{Fiederling,Schmidt02,Pareek04}. However, we  expect that the  spin current
which flows between leads 5 and 6 can generate a secondary
effect, the induction of a voltage difference in the top leg
between  leads 3 and 4. This  is in the same spirit as the
initially proposed set-up by Hirsch \cite{Hirsch99} but in a  far
simpler configuration without the  need of considering a bridged
conductor nor any unknown scattering mechanisms other than the
effects   of  disorder  which is  actually   small  in   this case
\cite{Sheng04,Nikolic04}.  We illustrate this  in Fig. 3 where we
show the   nonzero   voltage   difference  $\Delta
V_{34}(\lambda)-\Delta V_{34}(\lambda=0)$  as  a  function  of
Rashba coupling  for  different size systems, different meshes and
with  and without the additional leads 5 and 6. We  find the
increase of $\Delta V_{34}$ with the increase of Rashba coupling.
The induced voltage  variation is of  the order of  few $\mu$V for
$\lambda =60-80$ meV~nm and can  be easily measured.   We also
note that the inclusion or omissions of  leads 5 and 6 which in reality
cannot measure directly the spin Hall conductance  and do not  influence the
voltage difference; they  do  however  influence  the total
residual  voltage  background $\Delta  V_{34}(\lambda=0)$.  The
convergence of the results as a function of mesh size is
illustrated in Fig. 3 which show the results for the larger meshes
calculated.

We also note that at such disorder strength the results
do not depend on the disorder strength if we, say, double it.
This indicates two important facts: one, that the voltage induced is originating
from an intrinsic effect, and two, that the effect is not related to the constant
in-plane polarization induced by a current in a 2DES with Rashba coupling, the
Levitov effect \cite{Levitov}, since this effect is proportional to the mobility.

{\it Summary.} We have calculated, as a function of the Rashba SO
coupling strength $\lambda$, the voltage drop $\Delta  V_{34}$
that occurs in an H-shape sample in a response to a driving
dc-current between leads 1 and 2.  The voltage difference closely
follows the changes  of  the calculated SHE conductance, $G_{SH}$,
with $\lambda$. Moreover, $\Delta  V_{34}$
 increases  with the increase  of Rashba coupling  for constant
disorder  strength $W$  which is  a
clear  evidence  that  the observed  voltage signal  is
directly connected with the intrinsic SHE. Our work provides another
confirmation of the robustness of the intrinsic SHE against weak disorder
and shows the feasibility of detecting SHE signals through dc-transport
measurements in structures with realistic experimental parameters.
\section*{ACKNOWLEDGEMENTS}
We would like to thank N. A. Sinitsyn, K. Nomura, F. Sols, A. H.
MacDonald, Q. Niu, B. K. Nikoli\'c, and E. I. Rashba for stimulating discussions.
We also acknowledge financial support from  DFG (SFB410), 
the Czech Republic GACR (202/02/0912), and the DARPA SpinS program.


\end{document}